\documentclass[11pt, oneside]{article}   	
\usepackage{geometry}                		
\usepackage{graphicx}
\usepackage{subfig}
\usepackage{xcolor}
\usepackage{wasysym}
\usepackage{textcomp}
\usepackage{authblk}
\usepackage[utf8]{inputenc}
\usepackage{tikz}

\title{GeMSE: A new Low-Background Facility for Meteorite and Material Screening}

\author[1]{M. v. Sivers\thanks{moritz.vonsivers@lhep.unibe.ch}}
\author[2,3]{B. A. Hofmann}
\author[2]{\AA. V. Ros\'{e}n}
\author[1]{M. Schumann}

\affil[1]{Albert Einstein Center for Fundamental Physics, University of Bern, 3012 Bern, Switzerland}
\affil[2]{Institute of Geological Sciences, University of Bern, 3012 Bern, Switzerland}
\affil[3]{Natural History Museum Bern, 3005 Bern, Switzerland}

\date{}

\newcommand\copyrighttext{%
  \footnotesize Copyright 2015 American Institute of Physics. This article may be downloaded for personal use only. Any other use requires prior permission of the author and the American Institute of Physics.}
\newcommand\copyrightnotice{%
\begin{tikzpicture}[remember picture,overlay]
\node[anchor=south,yshift=1pt] at (current page.south) {\fbox{\parbox{\dimexpr\textwidth-\fboxsep-\fboxrule\relax}{\copyrighttext}}};
\end{tikzpicture}%
}

\begin{document}
\copyrightnotice

\maketitle

\begin{abstract}
We are currently setting up a facility for low-background gamma-ray spectrometry based on a HPGe detector. It is dedicated to material screening for the XENON and DARWIN dark matter projects as well as to the characterization of meteorites.
The detector will be installed in a medium depth ($\sim$620 m.w.e.) underground laboratory in Switzerland with several layers of shielding and an active muon-veto.
The GeMSE facility will be operational by fall 2015 with an expected background rate of $\sim$250 counts/day (100-2700 keV).
\end{abstract}


\section{Introduction}

\begin{figure}[htbp]
\begin{minipage}[t]{0.35\textwidth }
\centering
\vspace{0pt}
\includegraphics[height=0.2 \textheight]{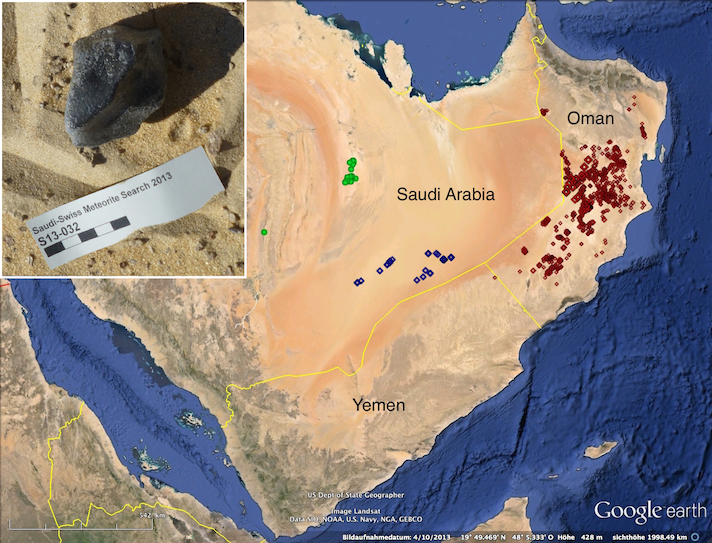}
\end{minipage}\hfill
\begin{minipage}[t]{.65\textwidth}
\centering
\vspace{70pt}
\begin{tiny}
\begin{tabular}{cccccc}
\hline
 \textbf{Isotope} &  \textbf{Half-life (y)} & \multicolumn{4}{c}{\textbf{Activity (mBq/kg)}} \\ 
 &  & \textbf{At fall} & \textbf{After 0.3\,y} & \textbf{After 20\,y} &\textbf{After 200\,y} \\
\hline
$^{22}$Na & 2.6\,y &1500 & 1385 & 7.3 & $\sim$10$^{-20}$ \\
$^{60}$Co & 5.3\,y &16.7 & 16.1 & 1.2 & $\sim$10$^{-10}$ \\
$^{44}$Ti & 63\,y & 16.7 & 16.6 & 13.4 & 1.8 \\
\hline
\end{tabular}
\end{tiny}
\end{minipage}
\caption{Left: Map of the meteorites found by the project "Meteorite accumulations of Arabia" (colored markers), the inset shows a specific sample found in the desert of Saudi Arabia.\\ Right: Typical activities of some cosmogenic isotopes in meteorites with different terrestrial ages.}
\label{fig:map}

\end{figure}

The new facility for low-background $\gamma$-ray spectrometry, GeMSE (Germanium Material and Meteorite Screening Experiment), will be dedicated to the characterization of meteorites as well as the selection of radiopure materials needed for rare event searches. It will be operated by two groups (geology/meteorite research and astroparticle physics) in a common interdisciplinary project. \\
The detection and quantification of cosmogenic isotopes in meteorites by $\gamma$-ray spectrometry is a non-destructive analysis method that allows for the determination of their terrestrial age, in particular the detection of relatively recent \cite{bhandari08,buhl14} and very old \cite{evans92} falls.
Our main interest lies in the detection of rather short-lived cosmogenic isotopes to identify fresh falls, mainly studying meteorite samples from the Oman collection \cite{hofmann04,al-kathiri05} hosted at the Natural History Museum Bern. This is one of the largest collections of hot desert meteorites comprising $\sim$880 fall events (see Fig.~\ref{fig:map}). 
From the activity of $^{22}$Na (T$_{1/2}$=2.6\,y) or $^{60}$Co (T$_{1/2}$=5.3\,y) we will be able to identify falls dating back to approximately 20 years. Using $^{44}$Ti (T$_{1/2}$=63\,y) one can recognize meteorites fallen 100-200 years ago. As can be seen in Fig.~\ref{fig:map} the needed sensitivity is $\sim$1\,mBq/kg. One of our main goals is an estimation of the average fall rate during the past 100 years, and the relative proportion of young falls among the whole collection, by screening the most unweathered and thus youngest samples.
We also plan a detailed research program on a series of fragments recovered from the Twannberg iron meteorite in Switzerland \cite{hofmann09}. The production of cosmogenic isotopes in meteoroids is strongly dependent on shielding (depth), and thus the meteoroid's size \cite{leya00,ammon09}. Fragmentation of meteorites during fall events will thus yield samples with highly variable activities of cosmogenic isotopes, reflecting the production rates at given depths and pre-atmospheric radii.
Statistical data on long-lived $^{26}$Al (T$_{1/2}$=7.2$\times$10$^5$\,y) in a series of samples will help to constrain the fallen mass, which likely represents one of the largest fall events in Europe. \\
The background goals of the future dark matter experiments XENONnT \cite{aprile14,plante14} and DARWIN \cite{darwin01} demand very radiopure construction materials. Many astroparticle physics experiments, searching for rare events, have shown that it is necessary to carry out extensive screening campaigns until all components are identified \cite{leonard08,aprile11}. As an example, a possible future DARWIN detector using 20\,t of LXe will require the screening of $\sim$1000 PMTs as well as several batches of PTFE, copper and stainless steel. The typical activities of isotopes from the U/Th chains in these materials are in the mBq/kg range but can be as low as $\sim$20\,\textmu Bq/kg \cite{aprile11}. The GeMSE facility will complement other HPGe detectors available to the collaborations at LNGS (Italy) \cite{arpesella91,baudis11} and in Heidelberg (Germany) \cite{heusser13,budjas07}. Besides providing additional resources with a very competitive background level, our facility will have the advantage of fast accessibility from Switzerland, which can be very relevant for delicate or urgent samples.

\section{Detector and Shielding}
\begin{figure}[htbp]
\centering
\includegraphics[height=0.3 \textheight]{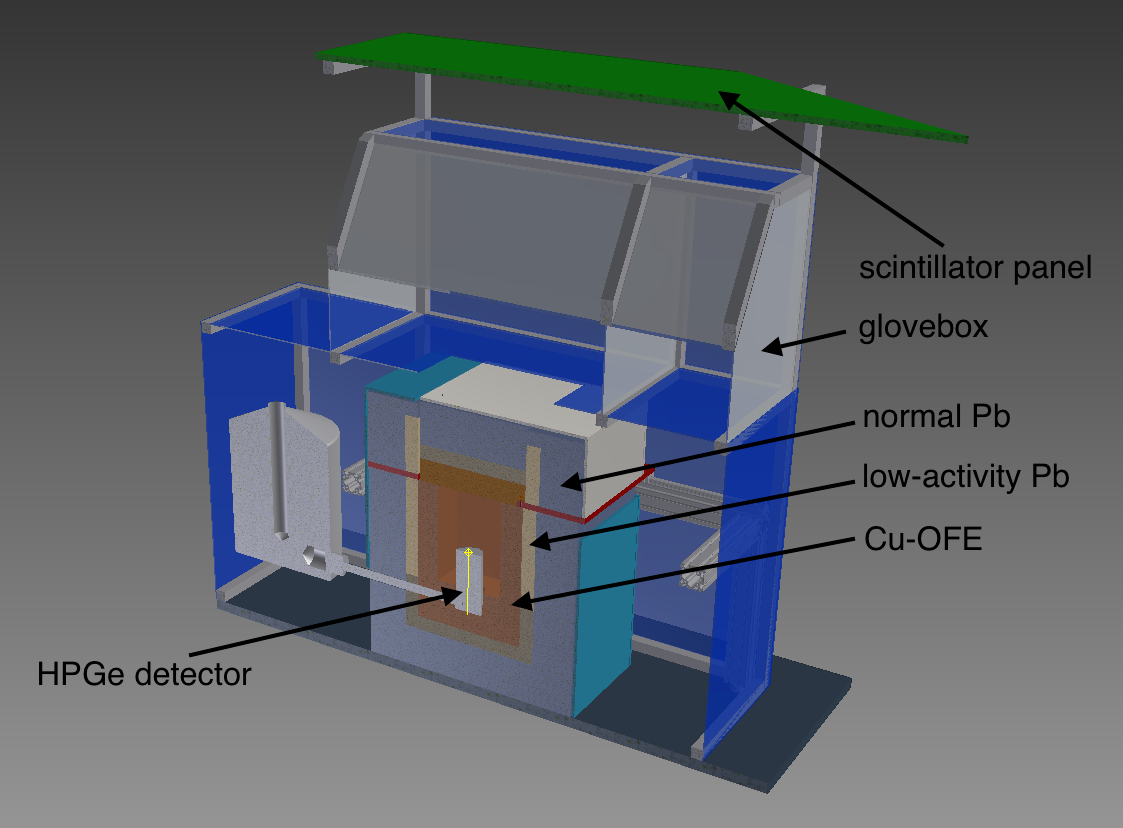}
\caption{Section view of the GeMSE setup. The HPGe detector will be surrounded by several layers of shielding and enclosed in a N$_2$ purged glovebox. A plastic scintillator panel on top serves as muon veto.}
\label{fig:setup}
\end{figure}

The detector of the GeMSE facility is a standard electrode, coaxial, p-type HPGe detector from Canberra with a relative detection efficiency of 107.7\%. The Ge crystal (\diameter=85\,mm, h=65\,mm) is embedded in a ultra-low background U-style cryostat made from Cu-OF. Our shielding design is schematically shown in Fig.~\ref{fig:setup}. The cavity for samples has a size of 24$\times$24$\times$35\,cm$^3$. From inside to outside the detector will be surrounded by 8\,cm of Cu-OFE, 5\,cm of low-activity Pb (7.2$\pm$0.5\,Bq/kg $^{210}$Pb) and 15\,cm of normal Pb ($\sim$200\,Bq/kg $^{210}$Pb). The whole shielding will be enclosed in a glovebox which is continuously purged with N$_2$ gas. Samples can be inserted with a lock system without introducing radon. In addition, a 120$\times$100\,cm$^2$ plastic scintillator panel on top is used as muon veto. The setup will be installed in the Vue-des-Alpes underground lab near Neuch\^{a}tel (Switzerland) \cite{gonin03}. It features a rock overburden of 235\,m corresponding to $\sim$620\,m.w.e. which reduces the muon flux by a factor of $\sim$1900. The lab is located in a highway tunnel $\sim$45\,min away from Bern and can therefore be easily accessed by car.

\section{Detector Characterization}
\begin{figure}[htbp]
\centering
\captionsetup{type=figure}
\subfloat[]{\includegraphics[height=0.25 \textheight]{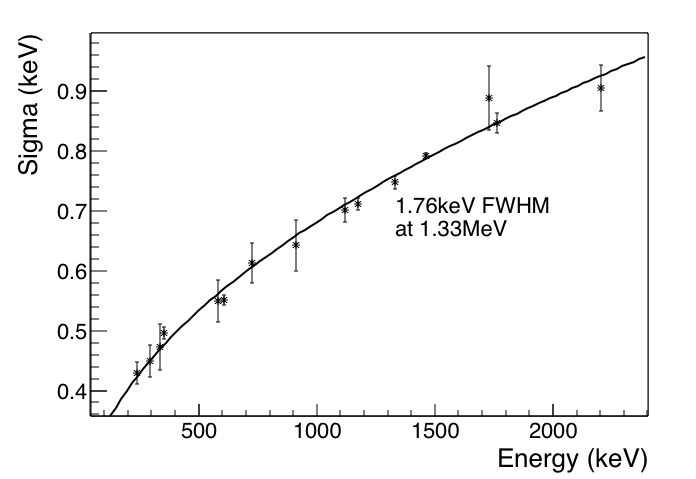}\label{fig:resolution}}
\subfloat[]{\includegraphics[height=0.25 \textheight]{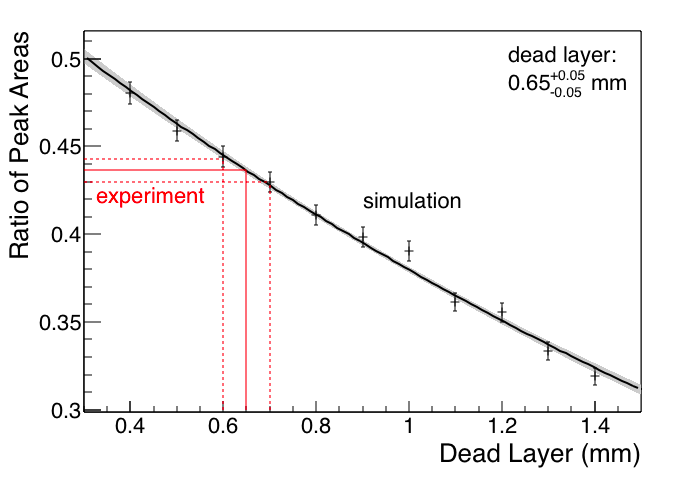}\label{fig:peakratio}}
\caption{(a) Energy resolution ($\sigma$) as a function of energy. The data points were fitted by the function $\sigma(E)=a\sqrt{E}+b$.
(b) The horizontal line shows the experimental value of the peak area ratio (81\,keV/356\,keV) of a $^{133}$Ba calibration (dashed lines indicate the statistical uncertainty). The data points show the peak ratio determined by a Geant4 simulation of the detector for dead layers of different thickness. The line shows a fit to the simulated data by an exponential function (shaded grey area indicates the 1-$\sigma$ error band). The vertical line indicates the determined dead layer thickness (dashed lines correspond to statistical uncertainty).}
\label{fig:characterization}
\end{figure}

Before the underground installation, we carried out first measurements to characterize our HPGe detector. 
Fig.~\ref{fig:characterization}\,\subref{fig:resolution} shows the energy resolution of the detector as a function of energy, measured with a shaping amplifier time constant of 6\,\textmu s. The resolution was determined for several peaks (mainly from U/Th chains) in a background spectrum measured without any shielding and an additional $^{60}$Co source. The resolution (FWHM) at the 1.33\,MeV peak of $^{60}$Co is 1.76\,keV (0.13\%).
To estimate the detection efficiency of a sample measurement it is important to know the thickness of the dead layer from the Li-diffused n+ contact. This thickness was determined by following the approach given in Ref.~\cite{agostini15}. A spectrum was recorded with a $^{133}$Ba source at a well defined distance (25\,cm) from the detector. From this measurement the ratio of the 81\,keV and 356\,keV peak areas was determined. This result was compared to the same ratio determined by a Geant4 \cite{agostinelli02} (version 9.6p03) simulation of the setup, performed with dead layers of different thickness. By matching the measured value to the simulation we get a dead layer thickness of (0.65$\pm$0.05)\,mm (see Fig.~\ref{fig:characterization}\subref{fig:peakratio}).  

\section{Background Simulation}
\begin{figure}
\begin{minipage}[t]{0.4\textwidth }
\centering
\vspace{0pt}
\includegraphics[height=0.22 \textheight]{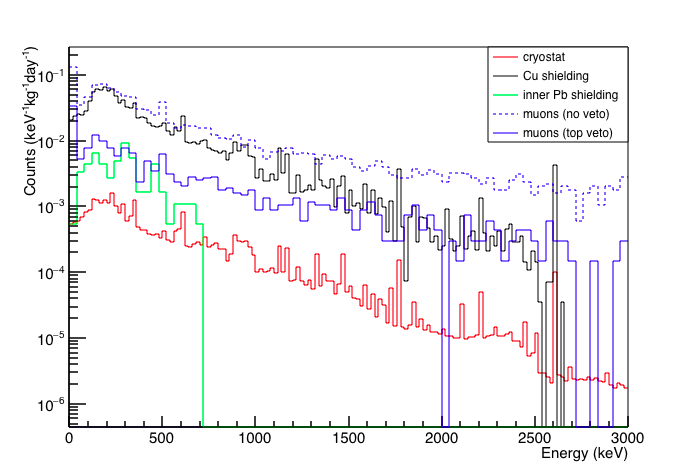}
\end{minipage}\hfill
\begin{minipage}[t]{.6\textwidth}
\centering
\vspace{58pt}
\begin{tiny}
\begin{tabular}{cccc}
\hline
 \textbf{Isotope} & \textbf{Cryostat} & \textbf{Cu shielding} & \textbf{Inner Pb shielding}   \\
\hline
$^{238}$U (\textmu Bq/kg) & 8 & 56 & - \\
$^{232}$Th (\textmu Bq/kg) & 4 & 27 & -\\
$^{40}$K  (\textmu Bq/kg) & 11 & 32 & - \\
$^{60}$Co  (\textmu Bq/kg) & 1.3 & 8 & - \\
$^{210}$Pb  (Bq/kg) & - & - & 7.2 \\
\hline
\end{tabular}
\end{tiny}
\end{minipage}
\caption{Left: Simulated background spectrum. Shown are the contributions from radioactive contaminations in the materials of the detector and shielding as well as that from cosmic ray muons with and without veto. \\ Right: Assumed activities of the different materials in the background simulation (note the different unit for $^{210}$Pb).}
\label{fig:simulated_background}
\end{figure}

The expected background for the GeMSE setup was estimated with a Geant4 (version 9.6p03) simulation. Initially, the simulation was also used to optimize the shielding design. The implemented geometry includes the detector with shielding, the muon veto panel and the laboratory cavern (including 2\,m of rock). We simulated the background from radioactivity in the cryostat, Cu shielding and the inner 5\,cm of Pb as well as that from cosmic ray muons. Figure \ref{fig:simulated_background} shows the values that we have assumed for the radioactivity of the cryostat and shielding components. For the cryostat and Cu shielding these were taken from the Gator screening facility \cite{baudis11} which uses the same type of HPGe detector and the same type of Cu in the shielding provided by the same supplier. The value of 7.2\,Bq/kg taken for the $^{210}$Pb contamination of the inner Pb layer was experimentally measured.
The flux, angular distribution and energy spectrum of the muons were calculated using empirical equations from \cite{grieder01} assuming a rock overburden of 623\,hg/cm$^2$.\\
The results of the simulation, namely the energy deposited in the Ge detector from each background source, are shown in Fig.~\ref{fig:simulated_background}. The dominant background component is from muons. It can be reduced by a factor of $\sim$6 using the top scintillator panel to reject all events in the Ge detector within a time window of 10\,\textmu s after an energy deposition $>$1\,MeV in the scintillator.
The total integrated background rate with (without) muon veto is 66 (128) counts/day (100-2700\,keV). Assuming a reduced veto efficiency and some additional background from other materials inside the cryostat and residual radon we estimate a realistic background rate of $\sim$250 counts/day, comparable to that of the Gator screening facility \cite{baudis11}.

\section{Summary and Conclusion}
The GeMSE facility will be a highly sensitive screening setup for low-background $\gamma$-ray spectrometry. The shielding is currently under construction and we expect the facility to be operational by fall 2015. With its underground location ($\sim$620\,m.w.e.), several layers of shielding enclosed in a N$_2$ purged glovebox and a plastic scintillator muon veto we expect a integrated background rate of $\sim$250 counts/day (100-2700\,keV). This is only a factor of $\sim$7 higher compared to the most sensitive screening facilities in the world \cite{heusser06}. GeMSE will help to answer important questions in meteoritics like the average fall rate or the pre-atmospheric size of the Twannberg meteorite. Furthermore, it will be used for the selection of radiopure materials for rare-event searches in astroparticle physics, such as the next generation dark matter experiments XENONnT and DARWIN.


\section*{Acknowledgements}
This project is supported by the interdisciplinary SNF grant number 152941. 
We thank A. Lazzaro at TU Munich for the screening measurement of the Pb sample and the mechanical workshop of LHEP Bern for their support.



\bibliographystyle{h-physrev}   

\bibliography{bibliography}

\end{document}